\documentclass[12pt]{article}
\usepackage{amssymb,amsmath,amsfonts,epsfig,graphics}

\def\e{{\rm e}}
\def\Re{{\rm Re}}
\def\Im{{\rm Im}}

\def\d{\partial}
\def\l{\left(}
\def\r{\right)}
\def\la{\langle}
\def\ra{\rangle}

\newcommand{\be}{\begin{equation}}
\newcommand{\ee}{\end{equation}}

\renewcommand{\ln}{\mathop{\rm ln}\nolimits}
\newcommand{\sm}[1]{{\scriptscriptstyle \rm #1}}

\newcommand{\bg}{\begin{gather}}
\newcommand{\eg}{\end{gather}}

\begin{document}
\title{On sgoldstino interpretation of HyperCP events}
\author{ D.~S.~Gorbunov\thanks{{\bf e-mail}: gorby@ms2.inr.ac.ru}, 
V.~A.~Rubakov\thanks{{\bf e-mail}: rubakov@ms2.inr.ac.ru}
\\
{\small{\em
Institute for Nuclear Research of the Russian Academy of Sciences, }}\\
{\small{\em
60th October Anniversary prospect 7a, Moscow 117312, Russia
}}}
\date{}
\maketitle
\begin{abstract}
We discuss possible sgoldstino interpretation of the 
observation, reported by HyperCP collaboration, of 
three $\Sigma^+\to p\mu^+\mu^-$ decay events with dimuon invariant 
mass $214.3$~MeV within detector resolution. We find that this
interpretation is consistent in models with (i) parity conservation in
sgoldstino interactions, (ii) pseudoscalar sgoldstino $P$ of mass
$214$~MeV and heavier scalar sgoldstino, and (iii) low supersymmetry 
breaking scale, $\sqrt{F} = 2.5$-$60$~TeV. In these models, 
sgoldstino can be observed in decays  $K^+\to\pi^+\pi^0 P$, 
$K_L\to\pi\pi P$ and
$K_S\to\pi\pi P$ at the branching ratios of at least of order
$10^{-12}$, $10^{-8}$ and $10^{-11}$, respectively. 
 The model can be probed also at $e^+e^-$
colliders in the channels $e^+e^-\to\gamma P$ and $e^+e^-\to e^+e^-P$,
where the cross section is likely to be in the range 1~pb - 5~ab.
\end{abstract}


\section{Introduction}
\label{sec:introduction}

Recently HyperCP collaboration has reported~\cite{HyperCP-PRL} the
first evidence for the decay $\Sigma^+\to p\mu^+\mu^-$ with branching
ratio at the level of $10^{-7}-10^{-8}$. The central value of the
measured branching ratio is larger than most Standard Model (SM)
estimates, although discrepancy with SM {\it in the overall branching
ratio} may not be dramatic in view of large uncertainties in the
hadronic matrix elements~\cite{He:2005yn}. The most remarkable property of
the HyperCP events is that dimuon masses of all three of them are
equal within the resolution of the detector. This
suggests~\cite{HyperCP-PRL} that the observed $\mu^+\mu^-$ pairs are
decay products of a new neutral particle $X$ born in the hyperon decay
\begin{equation}
\label{anomalous-decay}
\Sigma\longrightarrow p+X\;,~~~X\longrightarrow \mu^+\mu^-\;,
\end{equation}
with branching ratio
\begin{equation}
\label{anomalous-branching}
{\rm Br}(\Sigma\to pX(X\to\mu^+\mu^-))=[3.1^{+2.4}_{-1.9}({\rm stat}) 
\pm 1.5(\rm syst)]\times 10^{-8}\;,
\end{equation}
and mass of the intermediate neutral state $X$ equal to 
\begin{equation}
\label{X-mass}
m_X=214.3\pm0.5~{\rm MeV}\;.  
\end{equation}
As a phenomenologically viable explanation, supersymmetric models with
light sgoldstino and parity conservation~\cite{Gorbunov:2000cz} have
been proposed~\cite{HyperCP-PRL,PRESS-RELEASE}, the state $X$ being
interpreted as spinless particle $P$ (sgoldstino) --- one of the
superpartners of goldstino. 

This interpretation is further discussed in this paper.  Our
purpose is to estimate the relevant range of model parameters and to
discuss other experiments sensitive to such models. 

From HyperCP results we will estimate the strength of sgoldstino
coupling to muons, responsible for the sgoldstino decay
$X\to\mu^+\mu^-$, and sgoldstino flavor violating coupling to $d$- and
$s$-quarks, responsible for the hyperon decay~\eqref{anomalous-decay}.
In particular, we will find that the sgoldstino explanation of HyperCP
events implies the scale of supersymmetry breaking $\sqrt{F}$ within
the range 2.5-60~TeV (depending on hierarchy between MSSM soft terms
$M_{soft}$ and $\sqrt{F}$). The estimate of the flavor violating
coupling suggests that sgoldstino would show up in kaon physics. The
observed events cannot be attributed to {\it scalar } sgoldstino,
since in that case sgoldstino had to be already detected in two-body
kaon decays. It is the {\it pseudoscalar } sgoldstino $P$ that can
play a role of X-particle.

Pseudoscalar sgoldstino $P$, responsible for the HyperCP events, can
be observed in the decays $K^+\to\pi^+\pi^0 P$, $K_L\to\pi\pi P$ and
$K_S\to\pi\pi P$ at the branching ratios of at least of order
$10^{-12}$, $10^{-8}$ and $10^{-11}$, respectively.  With the inferred
value of $\sqrt{F}$, the model can be probed also at $e^+e^-$
colliders in the channel $e^+e^-\to\gamma P$ and $e^+e^-\to e^+e^-P$,
where the cross section is estimated as 1~pb - 5~ab depending on 
hierarchy of MSSM soft terms and the value of $\sqrt{F}$. Finally,
sgoldstino may be searched for via its resonant production in $e^+e^-$ 
collision at $\sqrt{s}=214.3$~MeV. 

The paper is organized as follows. In Sec.\ref{sec:models}
supersymmetric models with light sgoldstino are briefly discussed and
useful notations are introduced. The range of parameters, relevant for
the explanation of HyperCP events, is estimated in
Sec.\ref{sec:parameters}. In Sec.\ref{sec:tests} we study prospects of
various experiments in testing sgoldstino couplings.
Sec.\ref{sec:conclusion} contains concluding remarks.


\section{Models with light sgoldstino}
\label{sec:models}

Superpartners of goldstino --- longitudinal component of gravitino ---
may be fairly light. In a variety of models with low energy
supersymmetry they are lighter than a few GeV. Such pattern emerges in
a number of non-minimal supergravity models~\cite{ellis,no-scale} and
also in gauge mediation models if supersymmetry is broken via
non-trivial superpotential (see, e.g., Ref.~\cite{gmm} and references
therein). Here we merely assume that sgoldstino masses
are small, so that $\Sigma^+$-hyperon decay into proton and sgoldstino is 
kinematically allowed.  

Sgoldstinos couple to MSSM fields in the same way as
goldstino~\cite{bhat,comphep}, with corresponding coupling constants
being proportional to the ratios of MSSM soft terms (squark and
gaugino masses, trilinear coupling constants) to the supersymmetry
breaking parameter $F$ (which is of the order of squared scale of
supersymmetry breaking in the underlying theory). Constraints on
sgoldstino couplings may be translated into the limits on $\sqrt{F}$.
The most sensitive probes of sgoldstinos are flavor violating
processes~\cite{Gorbunov:2000th}, provided that flavor is violated in
squark and/or slepton sector. A sketch of the sgoldstino interactions,
experimental constraints on the models with light sgoldstino and 
relevant references can be found in Refs.~\cite{Gorbunov:2000th,comphep}.
                                               
A special class of supersymmetric extensions of the Standard Model in
which interactions of sgoldstinos with quarks and gluons conserve
parity but do not conserve quark flavor, has been considered in
Ref.~\cite{Gorbunov:2000cz}. Parity conservation in sgoldstino
interactions with quarks and gluons (as well as with leptons and
photons) may not be accidental.  As an example, it is natural in
theories with spontaneously broken left-right symmetry (see
Ref.~\cite{Gorbunov:2000cz} for details), which not only are
aesthetically appealing but also provide a solution~\cite{strongCP} to
the strong CP-problem.

It was found~\cite{Gorbunov:2000cz} that if the pseudoscalar
sgoldstino $P$ is light, $m_P <(m_K - 2 m_{\pi})$, and the scalar
sgoldstino is heavier, $m_S > (m_K - m_{\pi})$, an interesting place
for experimental searches is the poorly explored area of three-body
decays of kaons, $K^{0}_{S,L} \to \pi^+ \pi^- P$, $K^{0}_{S,L} \to
\pi^0 \pi^0 P$ and $K^{+} \to \pi^+ \pi^0 P$, with $P$ subsequently
decaying into $\gamma \gamma$, possibly $e^+ e^-$, $\mu^+ \mu^-$, or
flying away from the detector. Possible sgoldstino contribution to
hyperon physics has been considered in Ref.~\cite{QUARKS2004}. It was
shown that searches for hyperon decays into baryon and sgoldstino are
very sensitive to sgoldstino couplings in models with light
pseudoscalar sgoldstino, $m_P<(m_\Sigma-m_p)$, heavy scalar
sgoldstino, $m_S > (m_K - m_{\pi})$, and parity conservation. In what
follows we will consider models with parity conservation in sgoldstino
interactions, and make a remark on the parity-violating models in 
an appropriate place.

We are interested in trilinear couplings between sgoldstino and SM
fields, which can be derived by making use of spurion technique (see
e.g., Refs.~\cite{Brignole:1996fn,comphep}). The coupling constants of
sgoldstino to $SU(3)_c\times SU(2)_{\sm W}\times U(1)_{\sm Y}$ gauge
bosons are proportional to the ratio of gaugino soft mass terms to
vacuum expectation value of the auxiliary component $F$ of the spurion
superfield, while the coupling constants of sgoldstino to SM fermions
$f$ are proportional to the product of fermion mass $m_f$ and the
ratio of trilinear soft term $A_f$ to $F$. In particular, relevant for
our study {\it flavor-blind} sgoldstino interactions are determined by the
following effective lagrangians~\cite{comphep}: for pseudoscalar
sgoldstino $P$ 
\begin{equation}
{\cal L}_{P\gamma\gamma}=\frac{M_{\gamma\gamma}}{4\sqrt{2}F}PF_{\mu\nu}
F_{\lambda\rho}\epsilon^{\mu\nu\lambda\rho}\;,~~
{\cal L}_{P\bar f f}=-i\frac{m_fA_f}{\sqrt{2}F}P\bar f \gamma_5 f\;,~~
\label{P-SM-couplings}
\end{equation}
and for scalar sgoldstino $S$
\begin{equation*}
{\cal L}_{S\gamma\gamma}=-\frac{M_{\gamma\gamma}}{2\sqrt{2}F}SF_{\mu\nu}
F^{\mu\nu}\;,~~
{\cal L}_{S\bar f f}=-\frac{m_fA_f}{\sqrt{2}F}S\bar f f\;,~~
\end{equation*}
where $M_{\gamma\gamma}=M_1\cos^2\theta_{\sm W}+M_2\sin^2\theta_{\sm W}$
with $M_i$ being gaugino masses. It is worth noting that in many
models of supersymmetry breaking one has
$M_{\gamma\gamma},A_f\ll\sqrt{F}$, while the opposite case, 
$M_{\gamma\gamma},A_f\sim\sqrt{F}$, corresponds to unitarity limit.

Of special interest here are also the couplings of sgoldstinos to $d$-
and $s$-quarks, which {\it violate flavor}. In models with parity
conservation it is convenient to parameterize the corresponding
interaction terms as~\cite{Gorbunov:2000cz}
\begin{align}
\label{sgoldstino-sd}
   {\cal L}_{Pds} & = -P\cdot(h_{12}^{(D)} \cdot \bar{d}\, i\gamma^5 s
     + \mbox{h.c.})\;,~~~~\\
\label{sgoldstino-sd-1}
   {\cal L}_{Sds} & = -S\cdot(h_{12}^{(D)} \cdot \bar{d}  s
     + \mbox{h.c.})\;,
\end{align}
where 
\[
h_{12}^{(D)} = \frac{1}{\sqrt{2}}\frac{\tilde{m}_{D,12}^{(LR)2}}{F}
\]
and $\tilde{m}_{D,ij}^{(LR)2}$,
$i,j=1,2,3$ are left-right soft terms in the matrix of squared
masses of squarks. These interaction terms are assumed to be
responsible for the hyperon decay~\eqref{anomalous-decay}.


\section{Model parameters from HyperCP results}
\label{sec:parameters}

Couplings~\eqref{sgoldstino-sd} and \eqref{sgoldstino-sd-1} give rise
to the hyperon decays~\cite{QUARKS2004}
\[
\Sigma^+\to p P\;,
\]
and 
\[
\Sigma^+\to p S\;,
\]
respectively, whose rates can be estimated by making use of the matrix
elements
\begin{align*}
\la p|\bar{s}\gamma^\mu\gamma_5d|\Sigma^+\ra&=
B\cdot\bar{u}_p\gamma^\mu\gamma_5u_{\sm \Sigma}\;,\\
\la p|\bar{s}\gamma^\mu d|\Sigma^+\ra&=A\cdot\bar{u}_p\gamma^\mu 
u_{\sm\Sigma}\;,
\end{align*}
and neglecting the external momentum dependence of the form factors $A$ and 
$B$. The isospin symmetry implies that these constants are the
same as ones describing $\Sigma^-\to ne^-\nu$ decay, 
hence~\cite{Eidelman:2004wy}, 
\[
A=1\;,~~~~B=0.34\;.
\]
Then
\begin{align*}
\la p|\bar{s}\gamma_5d|\Sigma^+\ra&=
-B\frac{m_{\sm \Sigma}+m_p}{m_s}\cdot\bar{u}_p\gamma_5 u_{\sm \Sigma}
\;,\\
\la p|\bar{s}d|\Sigma^+\ra&=
A\frac{m_{\sm \Sigma}-m_p}{m_s}\cdot\bar{u}_p u_{\sm \Sigma}
\;,
\end{align*}
where $m_s$, $m_\Sigma$ and $m_p$ are masses of $s$-quark,
$\Sigma$-hyperon and proton, respectively. These amplitudes yield the
hyperon decay rates~\cite{QUARKS2004} 
\begin{align}
\label{sigma-pP-rate}
\Gamma_{\Sigma^+\to p P}&=\frac{|h_{12}^{(D)}|^2|B|^2}{8\pi}
\frac{\l m_{\sm \Sigma}+m_p\r^2}{m_s^2}
\l\l 1-\frac{m_p}{m_\Sigma}\r^2-\frac{m_P^2}{m_\Sigma^2} \r\cdot q_P\;,
\\
\label{sigma-pS-rate}
\Gamma_{\Sigma^+\to p S}&=\frac{|h_{12}^{(D)}|^2|A|^2}{8\pi}
\frac{\l m_{\sm \Sigma}-m_p\r^2}{m_s^2}
\l\l 1+\frac{m_p}{m_\Sigma}\r^2-\frac{m_S^2}{m_\Sigma^2} \r\cdot q_S\;,
\end{align}    
where 
\[
q_X=\frac{1}{2m_\Sigma}\sqrt{\l\l m_\Sigma+m_X\r^2-m_p^2 
\r\l\l m_\Sigma-m_X\r^2-m_p^2 \r}\;,~~~X=S,~P\;.
\]

Let us turn to the results of HyperCP collaboration~\cite{HyperCP-PRL}. 
Considering the central values of
Eqs.~\eqref{anomalous-branching}, \eqref{X-mass} as a reference point,
one obtains from Eqs.~\eqref{sigma-pP-rate}, \eqref{sigma-pS-rate} 
\begin{align}
\label{Pds-constant}
|h_{12}^{(D)}|{\rm Br}^{1/2}(P\to\mu^+\mu^-)&=3.8\cdot10^{-10}\;,\\
\nonumber
|h_{12}^{(D)}|{\rm Br}^{1/2}(S\to\mu^+\mu^-)&=6.8\cdot10^{-11}\;,
\end{align}
for pseudoscalar and scalar sgoldstino, respectively.  In models where
scalar sgoldstino $S$ is light, the strongest limit on \linebreak
$|h_{12}^{(D)}|^2{\rm Br}(S\to\mu^+\mu^-)$ comes from searches for
two-body charged kaon decays\footnote{The bound \eqref{5*} is
independent of the phase of $h_{12}^{(D)}$. Even stronger bound
\cite{Bellantoni:2005ca} is obtained from the search for $K_L\to\pi
S(S\to\mu^+\mu^-)$ decay \cite{Alavi-Harati:2000hs}. The latter bound
is valid, however, for ${\rm Re}[h^{(D)}_{12}]\simeq |h^{(D)}_{12}|$
only, and does not apply in the case of ${\rm Re}[h^{(D)}_{12}]=0$.} 
with subsequent sgoldstino decay into $\mu^+\mu^-$
(see Ref.~\cite{Gorbunov:2000th} for details),
\begin{equation}
\label{5*}
|h_{12}^{(D)}|{\rm Br}^{1/2}(S\to\mu^+\mu^-)<6\cdot10^{-12}\;.
\end{equation}
Thus, hyperon decay into proton and sgoldstino is highly suppressed in
these models, and the anomalous events observed in HyperCP experiment
cannot be attributed to light scalar sgoldstino.  In models with light
pseudoscalar sgoldstino and {\it parity violating} sgoldstino-quark
couplings the bound similar to \eqref{5*} applies to
$|h_{12}^{(D)}|^2{\rm Br}(P\to\mu^+\mu^-)$; this bound excludes
sgoldstino explanation of HyperCP result in those models.  The
remaining possibility, which is the focus of this paper, is
pseudoscalar sgoldstino and models with parity conservation. Then, the
strongest constraints~\cite{Gorbunov:2000cz} on $h_{12}^{(D)}$ come from the
study of $K^0-\bar K^0$ system and searches for $K\to \pi\pi P$
decays. The corresponding limits on $h^{(D)}_{12}$ are well above the
value~\eqref{Pds-constant}, hence the HyperCP events can indeed be
explained by pseudoscalar sgoldstino.

HyperCP collaboration did not present any estimate of $X$-particle
life-time, $\tau_{\sm X}$, since the opening angles of dimuons are
small and vertex resolution along the beam axis is rather poor. Using
reasonable estimates for the HyperCP decay vertex resolution (about 2
meters, see Ref.~\cite{Burnstein:2004uk}) and measured
$\gamma$-factors of the muons of anomalous events (about 270), 
one can place the upper limit 
\begin{equation}
\label{X-lifetime}
\tau_X\lesssim 2.5\cdot10^{-11}~{\rm s}\;.
\end{equation}

We note in passing that the above considerations in this section apply
to any pseudoscalar and scalar particles with couplings 
\eqref{sgoldstino-sd} and \eqref{sgoldstino-sd-1}.

For sgoldstino of mass~\eqref{X-mass}, only decay channels into
photons, $e^+e^-$, $\mu^+\mu^-$ and, possibly, gravitinos, contribute 
to the sgoldstino total width. Since sgoldstino-fermion 
couplings~\eqref{P-SM-couplings} are
proportional to the fermion mas\-ses, the
contribution of $e^+e^-$ is negligible (we will comment on the
possible opposite case in Sec.\ref{sec:tests}). The contribution of
invisible mode (decay into gravitinos), which is suppressed by
$m_P^2/M_{soft}^2$ (see e.g. Ref.~\cite{Gorbunov:2000th} for details)
is also negligible. Thus, sgoldstino width $\Gamma_P$ is saturated by
two contributions,
\begin{equation}
\label{dominant-decay-modes}
\Gamma(P\to\gamma\gamma)= 
\frac{m_P^3M_{\gamma\gamma}^2}{32\pi F^2}\;,~~~
\Gamma(P\to \mu\bar{\mu})=\frac{m_Pm_\mu^2A_\mu^2}{16\pi F^2}\l
1-\frac{4m_\mu^2}{m_P^2}\r^{1/2}\;.
\end{equation}
At $M_{\gamma\gamma}\sim A_\mu$ the latter contribution is suppressed
by an order of magnitude mainly due to the phase space factor. Thus,
without strong hierarchy between MSSM soft terms, sgoldstino of
mass~\eqref{X-mass} decays predominantly into two
photons. Unfortunately, HyperCP had no $\gamma$-detector, that
prevented the cross check of the sgoldstino explanation of the
anomalous events.\footnote{Note that coupling \eqref{sgoldstino-sd}
yields effective sgoldstino-neutron-$\Lambda$ interaction, but
sgoldstino with $M_P=213.4$~MeV is too heavy to be searched for in
$\Lambda$-hyperon decay. On the other hand, kinematically allowed is
the decay $\Omega^-\to\Xi^-+P$.}

What can be estimated from Eqs.~\eqref{X-lifetime} and
\eqref{dominant-decay-modes} is the upper limit on the scale of
supersymmetry breaking $\sqrt{F}$. At $M_{\gamma\gamma}=A_\mu=100$~GeV
one has $\sqrt{F}\simeq2.5$~TeV, while in the the unitarity limit,
$M_{\gamma\gamma}\sim A_\mu\sim\sqrt{F}$, one finds
$\sqrt{F}\simeq60$~TeV. Hence, sgoldstino explanation of HyperCP
events definitely implies {\it low energy scale of supersymmetry
breaking}.

We note that Eqs.~\eqref{dominant-decay-modes} also imply a
{\it lower bound} on sgoldstino lifetime. The shortest lifetime occurs
in the unitarity limit $M_{\gamma\gamma}\sim\sqrt{F}$ and
for $\sqrt{F}$ saturating the experimental bound
$\sqrt{F}\sim500$~GeV. One has 
\begin{equation}
\label{7+}
\tau_P\gtrsim 1.7\cdot10^{-15}~{\rm s}\;
\end{equation}
for $M_P=214.3$~MeV. 


\section{Possible experimental tests of sgoldstino evidence}
\label{sec:tests}

There are two independent direct tests of the sgoldstino explanation of
HyperCP results. One is related to sgoldstino flavor violating
coupling and another one is related to sgoldstino flavor-blind
couplings, whose strength has been estimated in the previous section. 

Sgoldstino-$d$-$s$ coupling gives rise to rare three-body kaon decays
into two pions and sgoldstino, whose branching ratios can be estimated by
making use of amplitudes presented in Appendix A with
$h_{12}^{(D)}$ given by Eq.~\eqref{Pds-constant}. The results for
neutral kaons depend on the possible hierarchy between real and
imaginary parts of $h^{(D)}_{12}$, which cannot be extracted from HyperCP
data. One obtains
\begin{equation}
\label{Ka}
{\rm Br}(K^+\to\pi^+\pi^0 P(P\to\mu^+\mu^-))
=1.2\cdot10^{-12}\;,
~~{\rm any}~{\rm Re}[h^{(D)}_{12}]/{\rm Im}[h^{(D)}_{12}]\;.
\end{equation}
For ${\rm Re}[h^{(D)}_{12}]\simeq0$, decays of $K_L^0$ are 
suppressed, 
\begin{equation}
{\rm Br}(K_L^0\to\pi^+\pi^- P(P\to\mu^+\mu^-))\simeq3.0\cdot10^{-13}\;, 
~~{\rm Re}[h^{(D)}_{12}]=0\;,
\end{equation}
while in the opposite case
\begin{align}
\label{Kb}
{\rm Br}(K_L^0\to\pi^+\pi^- P(P\to\mu^+\mu^-))&\simeq2.4\cdot10^{-9}\;, 
~~{\rm Re}[h^{(D)}_{12}]\simeq |h^{(D)}_{12}|\;,\\
\label{Kc}
{\rm Br}(K_L^0\to\pi^0\pi^0 P(P\to\mu^+\mu^-))&\simeq1.2\cdot10^{-8}\;, 
~~{\rm Re}[h^{(D)}_{12}]\simeq |h^{(D)}_{12}|\;.
\end{align}
On the other hand, decays of $K_S^0$ are suppressed at ${\rm
Im}[h^{(D)}_{12}]\simeq0$, 
\begin{equation}
{\rm Br}(K_S^0\to\pi^+\pi^- P(P\to\mu^+\mu^-))
\simeq5.2\cdot10^{-16}\;,~~{\rm Im}[h^{(D)}_{12}]=0\;,
\end{equation}
while without this hierarchy one has  
\begin{align}
{\rm Br}(K_S^0\to\pi^+\pi^- P(P\to\mu^+\mu^-))
&\simeq4.1\cdot10^{-12}\;,~~{\rm Im}[h^{(D)}_{12}]\simeq |h^{(D)}_{12}|\;,\\
{\rm Br}(K_S^0\to\pi^0\pi^0 P(P\to\mu^+\mu^-))
&\simeq2.1\cdot10^{-11}\;,~~{\rm Im}[h^{(D)}_{12}]\simeq |h^{(D)}_{12}|\;. 
\label{Ke}
\end{align}
These results are obtained to the leading order in chiral perturbation
theory. One may expect that actual numbers may be 30\% to 50\%
larger. Indeed, experimental data on the process $K^+\to \pi\pi
e\nu$, which is quite similar to the decays we consider, are
noticeable larger than the leading order predictions of chiral
perturbation theory (see, e.g.,
Refs. \cite{Colangelo:2000jc,Pislak:2003sv}).

For similar modes with sgoldstino decaying into photons one gets the
same numbers multiplied by
$\Gamma(P\to\gamma\gamma)/\Gamma(P\to\mu^+\mu^-)$. Therefore, the
branching ratios of the decays $K\to\pi\pi P(P\to\gamma\gamma)$ may be
substantially higher\footnote{As an example, for $A_\mu\sim\alpha_2
M_{\gamma\gamma}$ (which is still a fairly natural possibility),
branching ratios of $K\to\pi\pi P(P\to\gamma\gamma)$ are larger than
given in \eqref{Ka}~--~\eqref{Ke} by a factor of order $10^4$.}. Note
that decays into final state with neutral pions have significantly
higher rates, as compared to charged pions, because of larger phase
space. Note also that the above estimates for the branching ratios of
the decays (\ref{Ka})~--~(\ref{Ke}) apply not only to sgoldstino, but
to any pseudoscalar particle explaining the HyperCP events.

\begin{figure}[htb!]
\begin{picture}(0,0)%
\includegraphics[width=0.5\textwidth]{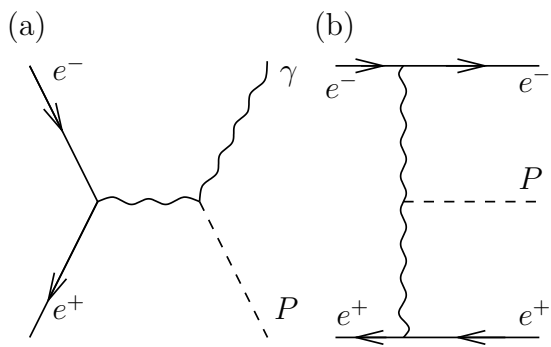}%
\end{picture}%
\setlength{\unitlength}{2819sp}%
\begingroup\makeatletter\ifx\SetFigFont\undefined%
\gdef\SetFigFont#1#2#3#4#5{%
  \reset@font\fontsize{#1}{#2pt}%
  \fontfamily{#3}\fontseries{#4}\fontshape{#5}%
  \selectfont}%
\fi\endgroup%
\begin{picture}(4544,2964)(579,-2243)
\put(800,100){$e^-$}
\put(800,-2000){$e^+$}
\put(2800,100){$\gamma$}
\put(2750,-2000){$P$}
\put(3200,-50){$e^-$}
\put(3300,-2050){$e^+$}
\put(4900,0){$e^-$}
\put(4900,-2050){$e^+$}
\put(4900,-850){$P$}
\put(400,500){(a)}
\put(3100,500){(b)}
\end{picture}
\caption{Amplitudes contributing to
sgoldstino production in $e^+e^-$ collisions; contributions to the
same final states, associated with sgoldstino-electron coupling, are
suppressed.
\label{collision}
}
\end{figure}

Flavor-blind sgoldstino-photon and sgoldstino-muon couplings can be
tes\-ted at $e^+e^-$ colliders. Sgoldstino coupling constants to leptons
are dimensionless and proportional to lepton masses, hence the most
promising processes of sgoldstino production involve sgoldstino-photon
coupling. Sgoldstino can be searched for in $e^+e^-\to\gamma P$ 
(fig.~\ref{collision}a) 
and $e^+e^-\to e^+e^- P$ 
(fig.~\ref{collision}b) 
with sgoldstino subsequently decaying into a
pair of photons or $\mu^+\mu^-$-pair. 

The cross section of $P\gamma$
production is
\begin{align}
\nonumber
\sigma_{P\gamma}&=\frac{\alpha}{12}
\frac{M_{\gamma\gamma}^2}{F^2}\l1-\frac{m_P^2}{s}\r^3\\
&=\frac{8\alpha\pi}{3}m_P^{-3}
\l1-\frac{m_P^2}{s}\r^3\Gamma(P\to\gamma\gamma)\;.
\label{sgoldstino-production}
\end{align}
The sgoldstino signals in different channels depend on the pattern of
sgoldstino branching ratios. If sgoldstino width is saturated by decay 
into photon pair, then, with account of \eqref{X-lifetime} and
\eqref{7+}, one finds from Eq.~\eqref{sgoldstino-production}
\begin{align*}
\sigma_{e^+e^-\to\gamma
P(P\to\gamma\gamma)}&=\frac{\sigma_{P\gamma}}{\Gamma(P\to\gamma\gamma)} 
\frac{1}{\tau_P}=64~{\rm ab}\div0.95~{\rm pb}\;,\\
\sigma_{e^+e^-\to\gamma 
P(P\to\mu^+\mu^-)}&=\frac{\sigma_{P\gamma}}{\Gamma(P\to\gamma\gamma)} 
\frac{1}{\tau_P}\frac{\Gamma(P\to\mu^+\mu^-)}{\Gamma(P\to\gamma\gamma)} 
=\l5.2~{\rm ab}\div77~{\rm fb}\r\cdot\frac{A_\mu^2}{M_{\gamma\gamma}^2}\;,
\end{align*}
where smaller and larger values refer to the lifetimes saturating
\eqref{X-lifetime} and \eqref{7+}, respectively. These formulae are
valid for $A_\mu/M_{\gamma\gamma}<3.5$, when the two-photon decay
mode is indeed dominant. 

In the opposite case, with dominant decay into $\mu^+\mu^-$, one
obtains 
\begin{align*}
&\sigma_{e^+e^-\to\gamma
P(P\to\gamma\gamma)}
=\frac{\sigma_{P\gamma}}{\Gamma(P\to\mu^+\mu^-)} 
\frac{1}{\tau_P}
\frac{\Gamma(P\to\gamma\gamma)}{\Gamma(P\to\mu^+\mu^-)}
=\l9.9~{\rm fb}\div 145~{\rm pb}\r
\cdot\frac{M_{\gamma\gamma}^4}{A_\mu^4}\;,\\
&\sigma_{e^+e^-\to\gamma 
P(P\to\mu^+\mu^-)}=\frac{\sigma_{P\gamma}}{\Gamma(P\to\mu^+\mu^-)} 
\frac{1}{\tau_P}  
=\l 0.8~{\rm fb}\div12~{\rm pb}\r
\cdot\frac{M_{\gamma\gamma}^2}{A_\mu^2}\;.
\end{align*}
The latter formulae are relevant for $M_{\gamma\gamma}/A_\mu<0.28$,
otherwise the assumption of dominant decay into $\mu^+\mu^-$ would not
be valid. 

The cross section of $e^+e^-P$ production, with the accuracy
$1/\ln\frac{s}{m_e^2}$, is~\cite{Eidelman:2004wy}
\begin{align}
\nonumber
\sigma_{e^+e^-P}&=8\alpha^2
\frac{\Gamma(P\to\gamma\gamma)}{m_P^3}\cdot
\l f(m_P^2/s)\cdot\ln^2\frac{s}{m_e^2}-\frac{1}{3}\ln^3\frac{s}{m_P^2}\r\;,\\
f(z)&=\l1+\frac{1}{2}z\r^2\ln\frac{1}{z}-\frac{1}{2}\l1-z\r\l3+z\r\;.
\label{sgoldstino-associated-production}
\end{align}
It grows with energy, becoming comparable with $\gamma P$ production
at $\sqrt{s}\simeq 0.5$~GeV and exceeding $\gamma P$ cross section by
a factor of 16 at $\sqrt{s}\simeq 10$~GeV. 

Another possibility of searching for light sgoldstino in
$e^+e^-$-annihilation is sgoldstino resonant production with the
beams' energy tuned to sgoldstino mass. In this case the cross section
of sgoldstino production exactly at the resonance peak is 
\begin{align}
\nonumber
\sigma_{e^+e^-\to P}&=\frac{2\pi^2}{m_P^2}\cdot {\rm Br}(P\to e^+e^-)=\\
\label{10*}
&=167~{\rm mb}\cdot 
{\rm Br}(P\to e^+e^-)\;.
\end{align}
If sgoldstino width is saturated by the decay into
$\mu^+\mu^-$, then at the peak
\[
\sigma_{e^+e^-\to P}=24~\mu{\rm b}\cdot\frac{A_e^2}{A_\mu^2}\;.
\]
If the $\gamma\gamma$ mode dominates over $\mu^+\mu^-$, then 
\[
\sigma_{e^+e^-\to P}=1.9~\mu{\rm b}\cdot\frac{A_e^2}{M_{\gamma\gamma}^2}\;.
\]
To estimate the number of events in realistic case one multiplies the 
above values by the product of collider luminosity and the ratio of
the sgoldstino width to the beam energy spread. The
latter is typically of order 
\[
\frac{\Delta p}{p}\equiv\eta\cdot 10^{-3}\;,
\]
with $\eta\sim1$. Thus, the suppression factor is
$\eta\cdot\l1.2\cdot10^{-10}-1.8\cdot10^{-6}\r$ for sgoldstino
lifetimes saturating the bounds \eqref{X-lifetime} and \eqref{7+},
respectively. Hence, in the case $A_e=A_\mu$, meaningful search for
resonant production of sgoldstino starts at integrated luminosity of
100~nb$^{-1}$, while integrated luminosity of up to 100~fb$^{-1}$ is
required to fully explore the parameter space.

Sgoldstino with energy $E_P$ in laboratory frame flies, before its
decay, a distance of about
\[
l_P=\frac{E_P}{m_P}\cdot(0.5~\mu{\rm m}-0.74~{\rm cm})\;. 
\]
Thus, there is an opportunity to resolve sgoldstino
vertices. 

Let us now discuss briefly exotic models with hierarchy $A_e\gg
A_\mu,M_{\gamma\gamma}$, where sgoldstino-electron coupling gets
enhanced as compared to models with the case $A_e\sim A_\mu\sim
M_{\gamma\gamma}$. First, $e^+e^-$ decay mode of sgoldstino never
becomes dominant: for sgoldstino with mass $m_P=214.3$~MeV the bound
\eqref{X-lifetime} on its lifetime gives unacceptably low scales of
supersymmetry breaking, $\sqrt{F}\lesssim200$~GeV even in unitarity
limit $A_e\simeq\sqrt{F}$. Second, upper limit on Br($P\to e^+e^-$) is
achieved for $A_e$ approaching unitarity limit $A_e\simeq\sqrt{F}$.
Then, in models where sgoldstino decays dominantly into photon pairs,
the upper bounds on Br($P\to e^+e^-$) are within $7\cdot
10^{-3}-1\cdot 10^{-6}$ for $\sqrt{F}=2.5-60$~TeV. In models with
sgoldstino decaying mostly into $\mu^+\mu^-$-pair, the upper bounds on
Br($P\to e^+e^-$) are between 0.7\% and $1.4\cdot 10^{-4}$ for
$\sqrt{F}=2.5-60$~TeV. 

Thus, search for sgoldstino in the decay channel $P\to e^+e^-$ is not
promising even for $A_e\gg A_\mu, M_{\gamma\gamma}$. However, the
cross section of the resonance production \eqref{10*} is considerably
enhanced for $A_e\gg A_\mu, M_{\gamma\gamma}$ as compared to the
conservative case $A_e\sim A_\mu\sim M_{\gamma\gamma}$, so in this
case sgoldstino would show up at relatively low integrated luminosity
(of order 1~nb$^{-1}$).


\section{Conclusions}
\label{sec:conclusion}
To conclude, we have identified the relevant region of parameter space
of models with light sgoldstino, which can be responsible for
anomalous events reported by HyperCP collaboration. We have estimated
sgoldstino couplings and have predicted rates of rare three-body kaon
decays and sgoldstino production in $e^+e^-$ collisions.  Sgoldstino
may be searched for in $\mu^+\mu^-$ and $\gamma\gamma$ decay channels, 
while $e^+e^-$ channel never dominates. Experimental tests of our
predictions will confirm or rule out sgoldstino explanation
of HyperCP results.

The special features of the supersymmetric models, capable of
explaining HyperCP results as evidence for sgoldstino, are: ({\it i})
low energy scale of supersymmetry breaking in the underlying theory
($\sqrt{F}\sim2.5-60$~TeV), ({\it ii}) parity conservation (left-right
entries in squark squared mass matrices make a hermitian matrix,
$\l\tilde m^{(LR)2}\r^\dagger=\tilde m^{(LR)2}$), ({\it iii}) tiny
flavor violation associated with $\tilde m^{(LR)2}_{{\sm D},12}$ entry
of squark squared mass matrix ($\tilde m^{(LR)}_{{\sm D},12}\lesssim
12~{\rm MeV}-1.4~{\rm GeV}$ for $\sqrt{F}=0.5-60$~TeV and sgoldstino
decaying mostly into $\mu^+\mu^-$; these bounds scale as ${\rm
Br}^{-1/4}(P\to\mu^+\mu^-)$ and become larger by an order of magnitude
with smaller but still natural $A_\mu\simeq \alpha_2\cdot
M_{\gamma\gamma}$ ).  The latter feature implies either fine-tuning,
or special flavor structure of MSSM soft terms preventing strong
flavor violation.

\vspace{3mm} 

We are indebted to D. Kaplan and M. Longo for extremely useful
correspondence. We thank M.~Danilov, V.~Kekelidze, Yu.~Kudenko, S.~Kulagin, 
V.~Obraztsov, P.~Pakhlov and A.~Skrinsky for stimulating discussions. 
This work was supported in part by the Russian
Foundation for Basic Research grant 05-02-17363 and by the grant of
the President of the Russian Federation NS-2184.2003.2. The work of
D.G. was also supported in part by the Russian Foundation for Basic
Research grant 04-02-17448, by the grant of the Russian Science
Support Foundation and by the fellowship of the "Dynasty" Foundation
(awarded by the Scientific board of ICFPM).

\vspace{3mm} 

{\bf Note added.} After the first version of this paper,
Refs.~\cite{He:2005we,Deshpande:2005mb} were posted in arXives.  We
agree with Refs.~\cite{He:2005we,Deshpande:2005mb} (as well as with
Ref.~\cite{Bellantoni:2005ca}) that scalar $X\equiv S$ in
(\ref{anomalous-decay}) is inconsistent with other data, while
pseudoscalar $X\equiv P$ is viable (it is pointed out in
Ref.~\cite{He:2005we} that axial-vector $X\equiv A$ is another
consistent possibility). The estimates given in Ref.~\cite{He:2005we},
their Eq.~(23), and in Ref.~\cite{Deshpande:2005mb}, Eq.~(8), are also
(almost) in agreement with our Eq.~\eqref{Pds-constant}.  Another
overlap with Ref.~\cite{He:2005we,Deshpande:2005mb} is the possibility
to search for $P$ in the decay $K_L^0 \to \pi \pi P~(P\to \mu^+\mu^-)$
and their results in the case of real sgoldstino-quarks coupling
constant are in numerical agreement with our estimates \eqref{Kb},\eqref{Kc}.

\section{Appendix A}

The part of chiral lagrangian relevant for $K_{L,S}\to \pi\pi P$
decays occurring due to coupling \eqref{sgoldstino-sd}, reads
\[
{\cal L}=\frac{f^2}{8}{\rm Tr}\left[ \d_\mu U^\dagger\d_\mu U\right] 
+ \frac{1}{4}B_0f^2{\rm Tr}\left[ \chi U^\dagger+\chi^\dagger U\right] 
\]
with $f=130$~MeV, $B_0=m_{\sm K}^2/(m_s+m_d)$ and 
\[
U=\e^{i\frac{2}{f}\Phi}\;,~~~~\Phi=
\begin{pmatrix}
\frac{1}{\sqrt{2}}\pi^0 & \pi^+ & 0 \\
\pi^- & -\frac{1}{\sqrt{2}}\pi^0 & K^0 \\
0 & \bar K^0 & 0 
\end{pmatrix}\;,
\]
\begin{align*}
\chi&=\hat M + i \hat H \cdot P\;,\\
2B_0\hat M & = diag(m_\pi^2,\;m_\pi^2,\;2m_{\sm K}^2-m_\pi^2)\;,\\
\hat H&=\begin{pmatrix} 0 & 0 & 0 \\
0 & 0 & h_{12}^{(D)} \\
0 & h_{12}^{(D)*} & 0
\end{pmatrix}\;,
\end{align*}
where isospin symmetry violation due to different masses of 
up- and down-quark is neglected. 

To the leading order in momenta one has the following interaction terms 
\begin{align*}
{\cal L}&={\cal L}^{kin}_{4}+{\cal L}^{pot}_{4}
+{\cal L}_{2P}+{\cal L}_{4P}\;,\\
{\cal L}^{kin}_{4}&=\frac{1}{3f^2}\l \bar K^0\d_\mu K^0 -K^0\d_\mu\bar K^0\r 
\l \pi^+\d_\mu\pi^--\pi^-\d_\mu\pi^+\r\\
&-\frac{1}{3f^2}
\l \bar K^0 \d_\mu\pi^- - \pi^- \d_\mu \bar K^0 \r 
\l K^0 \d_\mu\pi^+-\pi^+  \d_\mu K^0 \r \\
& -\frac{1}{6f^2}\l \bar K^0 \d_\mu\pi^0 - \pi^0 \d_\mu \bar K^0 \r 
\l K^0 \d_\mu\pi^0-\pi^0  \d_\mu K^0 \r \;,\\
{\cal L}^{pot}_{4}&=
\frac{m_{\sm K}^2+m_\pi^2}{3f^2} K^0 \bar K^0
\l
\pi^+ \pi^- +\frac{1}{2} \pi^0\pi^0
\r\;,\\
{\cal L}_{2P}&= 
B_0f \l K^0 h_{12}^{(D)*} + \bar K^0 h_{12}^{(D)} \r \cdot P\;, \\
{\cal L}_{4P}&= -\frac{2B_0}{3f}
\l
\bar K^0 h_{12}^{(D)}
+K^0 h_{12}^{(D)*}\r
\l 
\pi^+\pi^- + \frac{1}{2}\pi^0\pi^0
\r\cdot P\;.
\end{align*}

\begin{figure}
\begin{picture}(0,0)%
\includegraphics[width=\textwidth]{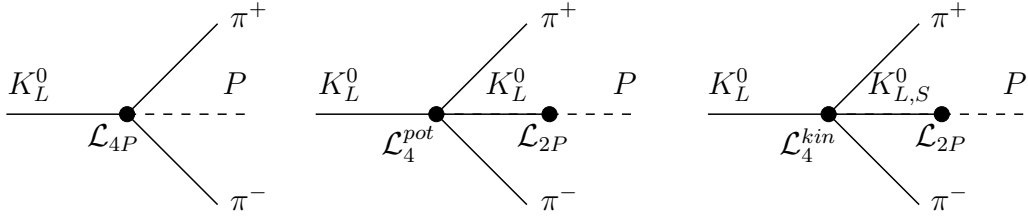}%
\end{picture}%
\setlength{\unitlength}{3947sp}%
\begingroup\makeatletter\ifx\SetFigFont\undefined%
\gdef\SetFigFont#1#2#3#4#5{%
  \reset@font\fontsize{#1}{#2pt}%
  \fontfamily{#3}\fontseries{#4}\fontshape{#5}%
  \selectfont}%
\fi\endgroup%
\begin{picture}(10244,1844)(279,-1283)
\put(300,-570){$K_L^0$}
\put(2250,-570){$K_L^0$}
\put(4700,-570){$K_L^0$}
\put(1650,-570){$P$}
\put(4100,-570){$P$}
\put(6550,-570){$P$}
\put(3300,-570){$K_L^0$}
\put(5700,-570){$K^0_{L,S}$}
\put(1700,-150){$\pi^+$}
\put(3650,-150){$\pi^+$}
\put(6100,-150){$\pi^+$}
\put(1700,-1300){$\pi^-$}
\put(3650,-1300){$\pi^-$}
\put(6100,-1300){$\pi^-$}
\put(800,-900){${\cal L}_{4P}$}
\put(2650,-950){${\cal L}_4^{pot}$}
\put(5150,-950){${\cal L}_4^{kin}$}
\put(3500,-900){${\cal L}_{2P}$}
\put(6000,-900){${\cal L}_{2P}$}
\end{picture}
\caption{Diagrams contributing to $K_L^0\to\pi^+\pi^-P$ decay. 
\label{Fig-kl-decays}
}
\end{figure}

One can rewrite the lagrangian in terms of mass eigenvalues 
of the neutral kaon system, 
\[
K^0_L\equiv \frac{1}{\sqrt{2}}\l \bar K_0+ K_0\r\;,~~~~
K^0_S\equiv \frac{i}{\sqrt{2}}\l \bar K_0-K_0\r\;.
\]
Then 
\begin{align*}
{\cal L}_{2P}&= 
\sqrt{2}B_0f \l K^0_L \Re h_{12}^{(D)} +  
K^0_S \Im h_{12}^{(D)} \r \cdot P\;, \\
{\cal L}_{4P}&= -\frac{2\sqrt{2}B_0}{3f}
\l
K^0_L \Re h_{12}^{(D)} + K^0_S \Im h_{12}^{(D)}
\r
\l 
\pi^+\pi^- + \frac{1}{2}\pi^0\pi^0
\r\cdot P\;,\\
{\cal L}^{pot}_{4}&=
\frac{m_{\sm K}^2+m_\pi^2}{6f^2}\l K^{02}_L +K^{02}_S\r
\l
\pi^+ \pi^- +\frac{1}{2} \pi^0\pi^0
\r\;,\\
{\cal L}^{kin}_{4}&= -\frac{1}{6f^2}
\l K^0_L \d_\mu\pi^- - \pi^- \d_\mu K^0_L \r 
\l K^0_L \d_\mu\pi^+-\pi^+  \d_\mu K^0_L \r \\
&-\frac{1}{6f^2}
\l  K^0_S \d_\mu\pi^- - \pi^- \d_\mu K^0_S \r 
\l K^0_S \d_\mu\pi^+-\pi^+  \d_\mu K^0_S \r \\
&-\frac{1}{12f^2}
\l  K^0_L \d_\mu\pi^0 - \pi^0 \d_\mu K^0_L \r 
\l K^0_L \d_\mu\pi^0-\pi^0  \d_\mu K^0_L \r \\
&-\frac{1}{12f^2}
\l  K^0_S \d_\mu\pi^0 - \pi^0 \d_\mu K^0_S \r 
\l K^0_S \d_\mu\pi^0-\pi^0  \d_\mu K^0_S \r \\
& +\frac{i}{2f^2}\l K^0_L \d_\mu K^0_S - K^0_S \d_\mu K^0_L \r 
\l \pi^+ \d_\mu\pi^- -\pi^-  \d_\mu \pi^+ \r \;.\\
\end{align*} 
Hence, there are three contributions to the amplitude of $K^0_L \to
\pi^+\pi^- P$ decay (see Fig. \ref{Fig-kl-decays}),

\begin{align*}
{\cal L}_{4P}\;:&\;\;\; -i\frac{2\sqrt{2}B_0}{3f}\Re h_{12}^{(D)}\\
{\cal L}^{pot}_{4}+{\cal L}_{2P}\;:&\; -i\frac{2\sqrt{2}B_0}{3f}\Re h_{12}^{(D)} 
\cdot \frac{m_{\sm K}^2+m_\pi^2}{2\l m_P^2-m_{\sm K}^2\r}\\
{\cal L}^{kin}_{4}+{\cal L}_{2P}\;:&\;\;\; 
i\frac{2\sqrt{2}B_0}{3f}\Re h_{12}^{(D)} 
\cdot \frac{m_P^2+m_{\sm K}^2+2m_\pi^2 
-3m_{\pi^+\pi^-}^2}{4\l m_P^2-m_{\sm K}^2\r}\\
&\;\;\;-\frac{2\sqrt{2}B_0}{f}
\Im h_{12}^{(D)}\frac{m^2_{P\pi^-}-m^2_{P\pi^+}}{4\l m_P^2-m_{\sm K}^2\r}\;,
\end{align*}
where $m_{ij}^2\equiv (p_i+p_j)^2$. 
The total amplitude to the leading order in momenta equals
\begin{align*}
{\cal M}\l K^0_L \to \pi^+\pi^- P \r&=
-i\frac{2\sqrt{2}B_0}{f}\Re h_{12}^{(D)} 
\cdot 
\frac{m_P^2-m_{\sm K}^2
+m_{\pi^+\pi^-}^2}{4\l m_P^2-m_{\sm K}^2\r}
\\
&-\frac{2\sqrt{2}B_0}{f}
\Im h_{12}^{(D)}\frac{m^2_{P\pi^-}-m^2_{P\pi^+}}{4\l m_P^2-m_{\sm K}^2\r}\;.
\end{align*}
The last term is the analogue of the exchange interaction. 

One has similar expressions for $K^0_L \to \pi^0\pi^0 P$ decay,
\begin{align*}
{\cal L}_{4P}\;:&\;\;\; -i\frac{2\sqrt{2}B_0}{3f}\Re h_{12}^{(D)}\\
{\cal L}^{pot}_{4}+{\cal L}_{2P}\;:&\;\;\; 
-i\frac{2\sqrt{2}B_0}{3f}\Re h_{12}^{(D)} 
\cdot \frac{m_{\sm K}^2+m_\pi^2}{2\l m_P^2-m_{\sm K}^2\r}\\
{\cal L}^{kin}_{4}+{\cal L}_{2P}\;:&\;\;\; 
i\frac{2\sqrt{2}B_0}{3f}\Re h_{12}^{(D)} 
\cdot \frac{m_P^2+m_{\sm K}^2+2m_\pi^2
-3m_{\pi^0\pi^0}^2}{4\l m_P^2-m_{\sm K}^2\r}
\end{align*} 
Note that the identity of final neutral pions implies 
the absence of the term proportional to the imaginary part of
$h_{12}^{(D)}$, unlike the case of 
$K^0_L \to \pi^+\pi^- P$ decay. 
The total amplitude to the leading order in
momenta equals
\[
{\cal M}\l K^0_L \to \pi^0\pi^0 P \r=
-i\frac{2\sqrt{2}B_0}{f}\Re h_{12}^{(D)} 
\cdot 
\frac{m_P^2-m_{\sm K}^2+m_{\pi^0\pi^0}^2}{4\l m_P^2-m_{\sm K}^2\r}\;.
\]

The same expressions are valid for $K^0_S$ with the only replacement
$\Re \leftrightarrow \Im$. 

\def\ijmp#1#2#3{{\it Int. Jour. Mod. Phys. }{\bf #1~} (19#2) #3}
\def\pl#1#2#3{{\it Phys. Lett. }{\bf B#1~} (19#2) #3}
\def\zp#1#2#3{{\it Z. Phys. }{\bf C#1~} (19#2) #3} \def\prl#1#2#3{{\it
Phys. Rev. Lett. }{\bf #1~} (19#2) #3} \def\rmp#1#2#3{{\it
Rev. Mod. Phys. }{\bf #1~} (19#2) #3} \def\prep#1#2#3{{\it
Phys. Rep. }{\bf #1~} (19#2) #3} \def\pr#1#2#3{{\it Phys. Rev. }{\bf
D#1~} (19#2) #3} \def\np#1#2#3{{\it Nucl. Phys. }{\bf B#1~} (19#2) #3}
\def\mpl#1#2#3{{\it Mod. Phys. Lett. }{\bf A#1~} (19#2) #3}
\def\arnps#1#2#3{{\it Annu. Rev. Nucl. Part. Sci. }{\bf #1~} (19#2)
#3} \def\sjnp#1#2#3{{\it Sov. J. Nucl. Phys. }{\bf #1~} (19#2) #3}
\def\jetp#1#2#3{{\it JETP Lett. }{\bf #1~} (19#2) #3}
\def\app#1#2#3{{\it Acta Phys. Polon. }{\bf #1~} (19#2) #3}
\def\rnc#1#2#3{{\it Riv. Nuovo Cim. }{\bf #1~} (19#2) #3}
\def\ap#1#2#3{{\it Ann. Phys. }{\bf #1~} (19#2) #3}
\def\ptp#1#2#3{{\it Prog. Theor. Phys. }{\bf #1~} (19#2) #3}
\def\spu#1#2#3{{\it Sov. Phys. Usp.}{\bf #1~} (19#2) #3}
\def\apj#1#2#3{{\it Ap. J.}{\bf #1~} (19#2) #3} \def\epj#1#2#3{{\it
Eur.\ Phys.\ J. }{\bf C#1~} (19#2) #3} \def\pu#1#2#3{{\it
Phys.-Usp. }{\bf #1~} (19#2) #3} \def\nc#1#2#3{{\it Nuovo Cim. }{\bf
A#1~} (19#2) #3}

\end{document}